**Title:** Entrepreneurship: what´s happening?

**Name:** Vítor João Pereira Domingues Martinho

**Affiliation:** Polytechnic Agricultural School of Viseu

**Address:**

Quinta da Alagoa - Estrada de Nelas

Ranhados

3500 - 606 VISEU

e-mail: vdmartinho@esav.ipv.pt

# Entrepreneurship: what´s happening?

**Abstract:**

Much has been said lately about entrepreneurship, so it seems important to leave here some personal analysis on this topic. The issues outlined here result from a work in about a year in which because a personal and professional obligations it was doing some research on these issues. This is an interesting topic that has not yet expired and on which there is much to research, do it is an area where there are many challenges.

# 1. Introduction

The literature on entrepreneurship is not really wide and to speak on this subject is also not very common in academic circles and even up in the midst of economic activities. Anyway, this is not a new subject, although it had a more prominent lately, much derived from the economic crisis and the need to maintain economies to grow to create jobs. It is known of the need for economies to grow about 2% per annum in order to create jobs. What then can end up clashing with environmental issues and lead to our "green foot" (which is necessary to remove and place in the environment for each of us to live) is growing. To reconcile economic growth with environmental sustainability is the major challenge of present and future brings us. Hence the appearance also of other concepts, such as sustainable development, namely economic growth without compromising the availability of resources for future generations.

There is a tendency to associate entrepreneurship to individuals, ie, entrepreneurship is because people are entrepreneurs. Moreover, some argued that this word derives from the French term "entrepreneur" that means the person who organizes, leads, operates and takes risks. But also, we associate entrepreneurship legal structures more like businesses, or even certain areas, as regions and countries. What makes perfect sense, particularly now, following EU policies, especially for rural areas, which speaks more to the need to make marketing space in order to promote so the set geographic areas through a brand space (Montemuro, Lafões, Caramulo, ...).

Hence, training in entrepreneurship is increasingly a concern of private institutions and also many public. Some argue the need to begin to start training in these areas in more original teachings, because in addition to awaken interest in these matters may be a new way of learning and to teach.

Usually a question that arises is whether entrepreneurship is born with the people or is acquired, or is something that is innate or is it possible to construct an entrepreneur. The views, of course, are divided on this issue, but it remains a good question.

For these reasons, we considered relevant to present the nineteen approaches as set out below (Martinho, 2011).

## 2. Another way of being

Entrepreneurs are increasingly central in our societies, because they are builders of economies, profit opportunities, accept the risks, play in the global market and when they reach the targets are wealthy.

The fact, currently, is that a job for life no longer exists, 80% of new jobs are created in a company with 10 or fewer workers, there needs to be adaptable, flexible and entrepreneurial, 45% of Europeans like to have your own business and 61% of Americans are ready to try to be entrepreneurs.

The evolution of societies leads parallel changes in the economy and education. A paradigm shift in education, with the various policies that have been adopted, and the "Bologna Process" is an example, has allowed students more perspective to the job market.

The effect, in the occidental world, is that in 1989 less than 5% of young people knew what was entrepreneurship, in 1999 64% considered it the first choice of career and ultimately the businesses started by young people increased four times, 50% of taxes come from small and micro-enterprise and in China created the "Wise Man Takes All" ("reality show" Chinese, in which competitors do their best to create your own business).

In these questions there is usually a gap between myth and reality. For example, there is the myth that mistakes have costs and the reality is that mistakes made early are profitable. Moreover, some say that we fail 100% of the shots that we do not try. Some argued that we should not make the management of failures, we should manage the costs of failures. Finally, Lee Iacocca (as Chairman of Chrysler Motors), said "Apply yourself. Get all the education You Need, by then, by God, of something. Do not just stand there, make it happen. "

In this regard it is worth seeing the videos on youtube of Steve Jobs, on a speech at Stanford University.

The economics have changed in many different levels and are now more organized in networks, have interdependent structures are based on the expectation of personal growth, they are led on the basis of charisma, have diverse work forces, looking to take advantage of the time, are customer focus and objective information. Consequently, education must become more focused on skills and less on content, the ownership of knowledge must be of the student, the leadership must be more of a mentor and fewer specialists, students need to be generating more ideas and minus liabilities, errors may be useful to learn and no longer feared and be an emphasis on practice.

### 3. Training models

There are several companies who have dedicated themselves to training in entrepreneurship (one of them is CG International (2010) which has several publications) based, some of them, one of best known models of training in these areas, as the model of CG International.

This model is supported by 15 years of experience in the development of entrepreneurship education in different parts of the world, is action-oriented and applies to personal and professional development. Does the entrepreneur as having a learning and development cycle. It focuses on personal development and the characteristics necessary to be an entrepreneur. At bottom, people are formed in order to make the most of their potential and thus can earn much money and get rich.

CGI's consultants have experience in a work done in over 65 countries and were directly or indirectly involved in the formation of 500 000 young entrepreneurs. Indeed, very recently held training at the Polytechnic Institute of Viseu, for teachers of various institutions of higher education, based on this model of CGI.

In this model the development cycle of the entrepreneur is divided into five phases: desire (I want to be an entrepreneur), generation of ideas (need an idea), identification of opportunities (there is an opportunity in that idea?), Planning and initiation. Of course, not all people can, even with such training, explore their potential and factors can be varied. Because if we were to take literally the theoretical literature on the characteristics of the entrepreneurs, those who could gather them all would be a superman or a superwoman. And here we refer to the old question, that is, entrepreneurs are born or build up.

The methods of generating ideas is about: copy, combine, solve problems, make something better, using leisure time, structuring capabilities, recycling, travel and looking for ideas, discussion groups, talking and listening, develop listings, find alternatives, improve something, daydreaming, etc... I would say that can generate good ideas is an art to which it is necessary to assemble a set of factors to propitiate.

The CGI model refers, among others, two methods of generating ideas, SCAMPER and WALT DISNEY.

The SCAMPER method whose name are initials of which should be linked to the generation of ideas, ie, Replace, Combine, Adapt, Modify, Put other uses, Eliminate and Reverse.

The method of Walt Disney is divided into three phases: the first phase of the Dreamer (Brainstorming is the phase in which no idea is bad), the second is the stage Designer (is where you make a selection of some ideas) and the last is the phase of Detail (where you choose the best idea).

Generally, entrepreneurs learn through the following phases: coaching, colleague support, experience, knowledge and planning. Coaching is a practical guide that gets used a lot on business and some people even use it to your personal life. In the background is a guardian, properly trained staff to guide people and / or professionally. Entrepreneurs must have personal qualities, attitudes, skills and information. After the foundation of entrepreneurship is the action. As such, an entrepreneur should be decided, should be aware that there are many factors around you that will destabilize and then must act (carry out an action, even without any information).

The techniques of teaching / learning model of CGI based mainly on practical activities, working in small groups, participants become owners of the learning process, who are themselves leaders and minimizes the time of traditional teaching in which the trainer minister theoretical concepts. It is curious that if you look at the legislation of the Bologna process much of what has been said about the CGI model ends up here and there, be found in this documentation. This seeks to focus the training to the level of European higher education on student and skills he should acquire, by changing the training paradigm for content for the training for skills.

In short, at this times, very derivative of the evolution of world economies and the current climate, those able to exploit optimally their capacity and make the right decisions in due time, it threatens to be a person of success.

**4. Organization of training**

The models of entrepreneurship, particularly those of CG International (developed and explained in CG International (2010) publications), usually also be applied in organizing the training.

In the CGI models are considered seven steps for the organization of training: introduction (it is a brief introduction to the topic and should focus on action to be held), clarity (to ensure that the student realized that the task will have to play) , action (the trainee perform the tasks that allow you to understand the context of learning), review (students should tell their stories and send in a simplified form what happened in the action), reflection (which is guided by the trainer should build bridges between action and matter), testing (works as an abstract) and widespread (connection between action and specific applications in different areas). In practice, these methodologies are not always easy to apply, much derived from the model of teaching and learning, yet we have at different levels of education, particularly in the early levels, because at the level of higher education, with the Process Bologna, the paradigm is changing, but here the situation is not easy because the same problem. In this regard it is important cite Peter Drucker, "The best way to predict the future is to create it.".

Plans for training the key points are: to support learners in their own learning process, facilitating an environment where trainees learn by doing, facilitate an approach to different modes of learning and heterogeneous group of learners, so that the exchange of experiences is greater. This is a formation that is not masses, but an individualized training, which then ends up colliding in some cases constraints of various kinds. Anyway, the philosophy is pertinent.

The dynamics of groups on these issues is crucial but must be prepared for the following stages: formation/incorporation, confusion/explosion, standardization, implementation and solution/suspension. At the stage of confusion/explosion appeals to the experience of the trainer is able to achieve the goals originally proposed.

**5. Entrepreneur for a day**

Usually the training on entrepreneurship culminates with the entrepreneurial activity for a day (and the training I did with the CG International was no exception), which seeks both to the students apply what they learned, contact with the reality of enforcing a idea in practice and to earn some money.

This activity should be: working in groups, plan the business for 3-4 hours, plan profit, non-recourse to lotteries or games, not developing illegal activities, be aware of existing business and not interfere with them, try to do anything different that has never been done before, not rely on charities as a method of sale and have fun. In this activity when you have large groups and people from diverse backgrounds is gratifying to witness the business ideas that are presented. The odd thing is that it often creates a learner motivation that leads to all to participate and enthusiastically.

In preparation of the business, entrepreneur for a day, students should seek to answer the following questions: what is the product/service you want to sell, who are the promoters of the business and customers, where is the business, the period in which will run the business, as customers know what is selling and how much will it cost business. In these questions some caution with the logistics, it is very easy to forget to ask permission to sell to municipalities in certain urban areas, the leaders of the shopping centers there to sell and so on. In most cases, just a verbal or telephone contact to request permission to occupy a certain space for 3-4 hours, sufficient time to implement a business for a day because more than that can become counterproductive. Incidentally, you can use this contact to just start doing business!

Once the activity is due to reflect on what happened using, among others, the method PNI (which was Positive, Negative and Indifferent). Positive about what was to be examined: what went right, what forces were used, stood up and was amazed to be successful at. In what was Negative must be taken into account: what needs to be improved, which could be different, which has disappointed and what would have been better. About Indifferent investigate: what surprised, what was thought to change and what is considered interesting. If planning for a particular action is crucial to the subsequent detailed analysis is essential to avoid repeating the same mistakes and try to improve in coming times.

### 6. From theory to practice

It is known the will of the Portuguese in creating their own business, but when it comes time to move from theory to practice, begin to emerge early hesitations. This matter has been subject to several sociological studies that have pointed to historical reasons, some recent and some not so.

It is imperative to change this trend, because we will lack the action, which sometimes has to do with bureaucracy, sometimes with the lack of associated services and still others with fear of making mistakes and yet with what other people might think.

The lack of associated services that enable putting into practice the idea of business is indeed a major constraint to business creation and entrepreneurial innovation. All of us had, already, innovative ideas which later foundered in the absence of services that help to make a good project, for example, or when there are, for lack of market competition, practice some prices misadjusted to the size of what is intended. Incidentally, I usually say that there are sectors where it is unclear why there is unemployment, because it's all to do.

The fear of failure is, as noted, a major barrier to innovation and entrepreneurship. In rare instances, in Portugal, who throughout his life story relates that he has created several businesses, or to mention that some of them lost money. But the way is like that, we need to implement the ideas that we think a priori, of course, are the best and when we fail it is unnecessary to be afraid to assume the mistakes, learn from them, avoid repeating them and pass for the implementation of another good idea. All this, with the costs of failures duly considered, it is preferable to inaction, to leave then and not risk. As already has been mentioned several times, resistance to frustration is the main characteristics of entrepreneurs.

In short, since the ideas are well planned (with good market research/marketing and with good financial studies), are the best partners and to gather good personal characteristics (honesty, and persistence), the chances of having success in the implementation good ideas are endless.

### 7. Self-employed

According to the European Commission (2010), 57% of the Portuguese like to have your own job, which will not only now, since already in 2000 were about 67% of the Portuguese wanted to work on their own. Moreover, in the first 15 European Union countries (excluding the CEECs - Central Europe and Eastern Europe), Portugal joined in this trend is only by Greece, Italy, Ireland and Spain. The scene in the new 10 CEECs is not much different, which means that the EU 25 (excluding Romania and Bulgaria which joined in 2007), about 50% prefer to be employed against the 37% of the United States of America. That shows the difference in terms of entrepreneurial spirit, these two worlds. If we think today, and socio-economic crisis which we live today, given this context, we can imagine the ability to recover these two realities, the European and North American. Although this indicator alone is not sufficient to draw conclusions, see the case of Portugal and Greece, which have a large percentage of the population wanting to create their own jobs and then did not materialize.

In the European Union to 25, men are more like to have your own job (50.2%), young people between 15 and 24 years (54.9%), the population also is studying (54.7%), those who have their own employment (72.9%) and those whose parents had their own job (50.6%).

The context referred to here for the European Union, ie the preference, on average, be employed, have causes in areas such as: the desire to have a guaranteed income, working hours fixed in social security, administrative difficulties, lack of funding for the actual job, not having worked on their own and fear of failure. Of all these reasons the most decisive in Portugal is the desire to have a guaranteed income and stable, as a little across the generality of the European Union.

### 8. Business creation

In the EU 25 (EU) 57% of the population is willing to create his own company, while the remaining would be willing to work solely on their own, without necessarily constitute an enterprise (European Commission, 2010). In the United States of America the picture is not very different in this aspect, ie around 54% of the population are interested in starting a business and invests in it. In this context, once again the Portuguese are above average, as are 71% who liked to create his own company.

In terms of socio-demographic factors in the EU are more men who liked having their own business (61.5%), young people aged between 15 and 24 years (65.3%), the population that is still study (67.8%), people from metropolitan areas (60.0%), those who have their own employment (59.5%) and those whose parents had their own job (62.5%).

The reasons that lead people to have their own business has to do with issues such as: personal independence, free choice of location and working hours, better income opportunities, business opportunities, favourable economic climate, lack of opportunities attractive jobs, a tradition of family or friends to create their own business.

In Portugal, like other European Union countries to 25, the main reason why people choose to create your own business is the personal independence and defining their own targets. If at first reason all countries are unanimous, in the second and third reason is not so, alternating between free time and place of work and the possibility of better performance.

If you look at these indicators alone, which has its value, as has already been said, there is satisfaction with the interest of young people to invest and take risks, but once again the rural areas fall far short of the desire to generate and implement new and innovative initiatives.

### 9. The image of entrepreneurs

In the European Union (EU) to 25 (without Bulgaria and Romania) in the 10 CEE countries (countries of Central and Eastern Europe and the newest EU members) and the United States of America (USA), approximately 88% population believes that entrepreneurs are creators of employment (with the U.S. to have a slight advantage and was around 89%), about 70% of people think are the basis for creating wealth, benefiting everyone (with the U.S. have Again, advantage and situate themselves in the 75% to 67% of the CEE countries), about 40% (roughly) is of the opinion that entrepreneurs only think in its own portfolio (52% in the CEECs and 24% in the U.S. ) and around 40% believes that exploit the people who work (57% in the CEECs and 26% in the U.S.). Interestingly, the two extremes in terms of opinion about the entrepreneurs in this set of countries. The U.S. with a more favourable opinion (result, eventually, a greater daily contact with the fruits of entrepreneurship) and the CEE countries with a more pessimistic view (the result probably of different economic strategies and guidelines followed in these countries, a story very recent).

In Portugal, 89% of Portuguese entrepreneurs consider the creators of jobs, 84% believe they are the basis of wealth creation, benefiting the entire population, 49% think they think only in their own enrichment and 52% are convinced that entrepreneurs' claims, only, to exploit the people who work.

In fact, both the EU-25, as U.S. and countries in CEE, invariably, people mostly think that entrepreneurs are job creators. In second place, these countries, it is considered entrepreneurs as creators of wealth, with the exception of Slovakia who regard them as exploiters of workers and Cyprus who regard them as selfish and concerned with the creation of his own fortune.

In social matters, in the EU25, the entrepreneurs are the basis of wealth creation, mostly for men, for people over 55 years, with training and with more than 20 years, metropolitan areas, with self-employment and whose parents had their own jobs. They are selfish and preoccupied with his own fortune, especially for women, for persons over 55 years, people in training and aged under 15 years of urban areas, workers on behalf of others whose parents were already employees. We are seen as creators of employment in most men, people 25 to 39 years, people with education and with more than 20 years, rural areas and with their own employment. They think they are explorers, especially men, people over 55 years, young people up to 15 years in training, and urban workers on behalf of others.

It's an interesting scenario and it shows as well as daily experiences ultimately influence the outlook of people on reality.

## 10. Tips for starting a business

The European Commission (2010) explains that one can start a business in 9 steps, namely: the conception of the idea, testing the idea and surround yourself the right team, development of the business plan, get the initial capital; formal constitution company, find the ideal place; definition of governing bodies and the recruitment of employees, and start the business.

The first major challenge is the conception of the idea. To help test the feasibility of an idea should seek to answer objectively the following questions: have the entrepreneurial profile, which the recipient of my product, the market needs of what I have to offer, what services they provide; which the benefits of my service, which my competition, how can I differentiate myself from my competition, what price will I charge for my services, which the initial investment that I will need, how will finance me, what is the best location for my company, the activity that will develop lacks any special licensing, which the capital that my company must have, there is some support for my work; choose as my partners, and how many ideal partners for my project.

In Portugal, the formal constitution should begin by choosing the ideal legal form for each company. From here you can pick up one of several centers of Business

Formalities to act on the following tasks: application for Certificate of Eligibility for Name or Designation of Collective Person; card application for Provisional Collective Person, marking deed, concluding deed, declaration of commencement of business; request the Commercial Register, publication in the Official Journal and entered in the National Register of Legal Persons; enrolment in Social Security, and application for entry in register of Commercial or Industrial.

Who meet all these stages, is capable of initiating activity of your company. At this point there should be a concern to verify that all details are operational to receive customers, from facilities, human resources and communication structures.

### 11. To avoid errors in setting up a business

Like everything in life, too, in the creation of companies is usually the small details and the small questions, often overlooked by seem harmless, that problems arise.

Therefore, according to the European Commission (2010), there are mistakes we should avoid in the creation of a company, such as miscalculating the market; underestimate the competition; invest prematurely; misjudge deadlines; misjudge the foreseeable profitability ; know the sector; customize too firm, not properly consider the legal issues, and conflict of ideologies between the partners.

Miscalculate the market may be the verdict of failure at the outset of the company, because it usually is not difficult to produce but to sell and, therefore, we must know the company's market and produce according to consumer needs that market.

Competition should always be respected and never devalued, and for that we must always be ahead of what the company´s potential competitors can offer and market.

Investing more than is necessary and sizing the company may have consequences that can be decisive at the very start, especially when the investment is made with excessive use of banking and payment of interest. It may happen that the turnover does not get to pay interest and amortization.

Calculate well the "timings" of business is crucial for success in order to avoid the market products so lagged and times unlikely to be consumed. It is also important to know the area well so you can steer, knowingly, the evolution of the company. Importantly, though, not all focus on one person, but working in a team, with diverse opinions of each, can only bring benefits to businesses.

## 12. Aspects to take into account when setting up a company

In this era of globalization and ease of movement of information, protecting the results of innovative businesses through Intellectual Property is critical (empresanahora, 2010). The Intellectual Property consists of Industrial Property and the Copyright and relates to human creations, including the level of inventions and artistic and literary works. With respect to the Intellectual Property with the provisions of its code are: patents, utility models, trademarks and other distinctive signs. Much of the national and European public policy, to which Portugal is subject, are all geared towards innovation, hence this issue is essential for any business wishing to succeed.

In the context of policies to simplify administration, creation of company and brand at the time nowadays is a reality that prevents economic agents from getting lost in the lengthy bureaucratic processes. The creation of enterprises through this initiative, in Portugal, is a simple process that includes the following phases: choosing a name and a social pre-approved pact for the company in any office "On the Spot Firm"; deposit the social capital in any bank; designate a Accountant; and submit a declaration of commencement of activity in the service of "On the Spot Firm" or the Finance Service. The choice of the mark, too, is relatively simple and can be made from a stock of firms and marks.

In terms of financial support for business creation there are several tools available, including the NSRF (National Strategic Reference Framework) and PRODER (Rural Development Programme). There are several entities on the other hand, who see support for the creation of new businesses, such as business opportunities. In this case, I refer to banks and financial, among others, firms "Business Angels".

Innovation and entrepreneurship are new challenges in a globalized era and all that concerns them should be taken seriously as a matter of survival of firms that compete in free markets and highly competitive.

### 13. The need to find a good idea

The most important element in both the EU-25 (EU) and the United States of America (USA), to create a business is to have a proper idea. Refer to as needed, about 84% of people in the European Union and about 80% in the United States of America. It follows, for similar percentages, the need for adequate financial means of support. About 60% of the population on average for the EU and the U.S., says that having good partners is essential, and finally about 55% in average population indicates how important the change in the family context, with the U.S. to present here a slight advantage over the EU unlike that in the background, possibly a reflection of family structure in the United States of America.

On this issue the Portuguese population has similar views to those of the European Union and the United States of America, ie: 87% consider the existence of an idea as an important element to create a business, 79% means the existence of financial support , 76% having good partners, 72% have a dissatisfaction with the current situation and 64% had been no change in the family context.

But not all EU countries are unanimous in considering the most important element for creating a business. Although most countries consider the existence of an idea as paramount, there are countries such as France, Slovakia, Cyprus and Latvia they consider most important to have funding available to support the initiatives.

Are women, youth and people of the metropolitan areas, the share of EU population that gives more importance to ensuring a good idea to create a viable business.

By analyzing these data we conclude that the existence of good ideas is key to moving forward with new initiatives. Hence, it may be here a little explanation for the difficulty in translating into practice the will to create their own employment manifested, for example, by population.

### 14. The education system and entrepreneurial attitudes

About 63% of the population of the United States of America (USA) believes that the education system helped in the development of entrepreneurial attitudes against 50% of the European Union (EU). More or less the same percentage of the population, in both parts of the world, believes that the educational system helped to better understand the importance of entrepreneurship in society. For the lower slices of the population (around 30%), but always with the United States of America to lead, people believe that the education system was able to arouse their interest to become entrepreneurs.

In Portugal, about 71% of people believe that education helped them to develop entrepreneurial attitudes, almost 75% believe that the school helped them understand the role of entrepreneurship in society and around 50% of the population is opinion that the education they received awakened their interest to become entrepreneurs.

In the European Union are the men and young, between 15 and 24 years, the ones that are of the opinion that the school helped them to develop entrepreneurial attitudes, and to clarify the importance of entrepreneurship in society.

Interestingly when people are asked if participating / participated in any courses related to entrepreneurship at the school, only about 13% of Americans answered yes to 34% of Europeans. What is not is curious, because as it turns out behind the Americans greatly value the school as a means of encouraging entrepreneurship and clarify and then not have a specific course for enterprise, contrary to what is happening with Europeans.

I think there are still many lessons to be drawn from these data and the need to constantly review these settings and make the adjustments necessary and appropriate.

**15. What prevents people from taking risks**

In the EU 25 (EU), 30% of the population is referred to the possibility of failure in business that prevents economic agents to take risks, 20% say it is the uncertainty of income, 18% the risk of losing control business, 8% uncertainty of employment and personal losses, and 9% of businesses that require too much time and energy. In the United States of America (USA) people's opinions are similar, although the Americans consider that the uncertainty over the income that most frightens the economic time to take risks. Concern in 25% of the people, against 23% for the second answer given with respect to the possibility of failure in business.

In Portugal 32% of the population says it is the possibility of failure that prevents people from taking risks, says that 22% is the uncertainty of income, 15% the risk of losing business, 8% uncertainty of employment, 11% risk of having personal losses and 3% that require much time and energy.

Comparing the evolution of views on this subject, from 2004 to 2007, the fear of bankruptcy was the largest growth in the European Union, while in the United States was the fear of losing ownership of enterprises.

Whereas investing in micro, small and medium-sized company that can create jobs and generate wealth more easily, since they are simpler to create niche markets that the economic dynamics are being opened, it is urgent to prepare people so they seize the business opportunities they arise.

From the data presented here are significant slices of the population who reported a set of fears that it prevents them from moving towards the creation of businesses.

It also admits that some of these issues can only be resolved with time, but there is much work to do that can pass through more effective strategic policies at various levels.

### 16. Innovation in agriculture

Agriculture in Portugal still has a long way to go, indeed I dare myself to say that almost everything is still to be done. So the opportunities are huge, because when we leave the current agricultural models existing in a significant portion of our territory, is based on a subsistence logic (very useful in other times), the possibility of wealth creation is significant.

The need does the man and it has been said that in Portugal we have never had much need to have a good agricultural sector as a result of several factors, but some people said which the discoveries had here a very decisive weight. Perhaps, these prophecies have some reason to be because when we consider that today the Israelis are the greatest experts in irrigation systems by having the need to conquer the desert terrain, and that the Dutchmen are specialists in drainage, for having conquered land to water, there may be some truth in these views.

Anyway, we have good examples of very successful agricultural business and innovative examples in the sector, even in terms of research, as the project developed with support from the Innovation Agency for the Development of superabsorbent polymers for forestry and combating desertification. This project is characterized by the development of new superabsorbent polymers for water retention in soil and application technologies of superabsorbent in combating desertification by forestry. The superabsorbent polymers were synthesized from monomers available for industrial production and the resulting hydrogels were tested for reversibility of the cycle: water absorption, transfer to the plant, re-absorption. The absence of leachable fractions, the resistance to hydrolysis and biodegradation were tested aspects of the project.

### 17. Innovation in agri-food sector

When we see the website of Innovation Agency we find another good example of innovation on the agro-food industry. In this case, I speak of a project on "Optimisation of quality and cost reduction in the distribution chain for fresh agricultural products".

This project has been completed but it is worth explaining it based on information provided by the Innovation Agency. The project's objective was to design a new methodology for chain manufacturer-distributor-consumer ensuring on the one hand the reduction of the costs and secondly the quality of fresh produce. The project had the following specific objectives: to apply the HACCP concept to the entire post-harvest chain, developing innovative packaging systems, using cutting-edge scientific knowledge, disseminate information to participants in the chain, especially farmers.

The targets set were achieved by implementing the following steps to six products - strawberries, peaches, kiwi, lettuce, tomato and carrot: preliminary analysis of the chain for each product; characterization of production conditions, characterization of products, optimizing the harvest; setting up a HACCP team and a HACCP plan; characterization of materials and packaging systems; development and implementation of packaging systems, disseminating information about the project.

In scientific terms, the most innovative results of this project relate to packaging systems optimized for the selected products. Another innovative feature is the set of actions that could begin to be implemented and which constitute an interesting change in the way they act and interact in a market effectively competitive producers. As elements placed at different points in the chain, all the consortium members identified in this project contributions to their areas of business, easily extensible to other stakeholders.

Stakeholders in this project were: the School of Biotechnology, Portuguese Catholic University; SONAE, SA; Institute for Agrarian Development in the Northern Region; Portucel, SA; Regional Direction of Agriculture of Entre Douro e Minho.

### 18. Fodder production alternatives in bovine milk

By passing again by the website of the Innovation Agency found another example of innovation in agriculture. The project that I speak on is "fodder for dairy alternatives."

This project has been completed but it is worth some considerations based on information provided by the Innovation Agency. With this project was possible answer to important aspects of adaptation level, productivity, nutritive value and silage from a wide range of legumes, on the other hand raised new issues and research opportunities. Some issues that need further analysis resulted in several theses of doctoral and masters, all directly involved with the subjects under study. This project was also behind the production of others that are somewhat complementary, in particular the use of silage for pigs, and of immature cereal silage for dairy cows, meanwhile submitted and already approved by the AGRO Programme - Measure 8.1 of the Ministry of Agriculture, Rural Development and Fisheries.

In concrete terms, has achieved an important set of information about alternative forages for dairy cows obtained in the region of Entre Douro e Minho, and set up the technology of culture and silage (particularly with regard to pre-drying, the length of the section, the use of additives, preparation of silage in big bales, and the mixture and complementarity of fodder). They were also successful the most visits to several farms in the region, which encouraged farmers to adopt new techniques.

This project had the participation of the following: Union Co-operative Milk Products of Entre-Douro e Minho and Trás-os-Montes; Institute for Agrarian Development in the Northern Region; Institute for Science and Technology Agriculture and Agri-Food; Regional Direction of Agriculture of Entre-Douro e Minho; and Agronomic Superior Institute.

**19. And if we did all okay**

Southern countries generally have the fame, and some say the advantage of being less organized, with all consequences arising from this. Following this, we can think about if those countries did an effort to do everything well.

Maybe it is worth imagining how it would be our day tomorrow if we got to our commitments on time, if we had a proper social behaviour (no scream in traffic), if we did our daily duties with perfection and if we made an effort to have positive thoughts.

Utopian, many are thinking, but maybe we would be able to produce much more and with much less effort (and there are countries where this is possible). We have policies for innovation and entrepreneurship fantastic, but if we do not have correct cultural context, we can hardly walk a couple of countries where the organization is very well planning.

Every time I analyze the policies of the European Union, so for twenty-seven member countries, I am worried and expectant as some countries will be able to compete with other nations.

Last week I was in a Portuguese University to attend the presentation of the Framework Programme for research and I could verify that the rules are too tight to some Portuguese institutions be able to approve projects in this area. I would say that increasingly will be able to make a difference who is able to perform tasks perfectly, contacting the right people and respect their commitments. Of course some will be thinking that I am again to be utopian, but if there are those who can accomplish this, we too will be able and obliged to do so if we want to enter in the international championship which is increasingly where the major financial supports are played.

**20. The era of bad mood**

A few years ago, I was still a novice in these labours of teaching and for a work to my classes at secondary school, I bought a technical-scientific journal to prepare an activity for my students in that time. It was an interesting magazine with very good articles and thought it worthwhile to buy it for use as a basis, some articles written there.

On the cover appeared a bold print a summary of an article that was titled more or less as follows: the era of the bad mood. Of course, this title caught my attention and I read the article. Reading more I identified a set of situations we find every day in many different sites and services (public or private). In short the idea is that we find in our day people willing evil and people well prepared, but we tend to flee from evil willing to do not spoil the day and at the end the well prepared remain with bad mood.

I write about this here, because I am convinced that this happens in many places and is a barrier to innovation and ultimately the development of the country. We have failed at different sites and services in the country, when we did not put people with the most appropriate profile in the right places. We still think that everyone has capabilities for everything and because that we have the difficulties we found every day.

So I see with good eyes, for example, the decision of consider in the public admissions the psychological tests, to try adjusting the profile to the functions. I hope that this would prevent many of the problems we face daily and which are a clear obstackle of the country's development.

## 21. Conclusions

Finally, I would say that entrepreneur is not anyone, but who can, because it is a task that requires a set of characteristics that in these times, of easy solutions, not abound. We must be willing to give up our comfort zone every day, which is incompatible with this idea that we watch these days that everything can be achieved without effort. Because this I think we live days relatively complicated, namely in Portugal.